\documentclass[preprint]{aastex}
\renewcommand\bv{\mbox{\boldmath $v$}}
\renewcommand\>{{\rangle}}
\newcommand\qe{{\tilde{q}}}

\newcommand\bnabla{\mbox{\boldmath $\nabla$}}
\newcommand\bk{\mbox{\boldmath $k$}}

\newcommand\bx{\mbox{\boldmath $x$}}
\newcommand\ex{\mbox{\boldmath $\hat{x}$}}
\newcommand\ey{\mbox{\boldmath $\hat{y}$}}
\newcommand\ez{\mbox{\boldmath $\hat{z}$}}
\newcommand\bO{\mbox{\boldmath $\Omega$}}
\newcommand\bxi{\mbox{\boldmath $\xi$}}
\newcommand\<{{\langle}}

\newcommand{\pdv}[2]{\frac{\partial#1}{\partial#2}}
\newcommand{\dv}[2]{\frac{d#1}{d#2}}
\newcommand{\be}{\begin{equation}}
\newcommand{\ee}{\end{equation}}
\newcommand{\Ri}{{\rm Ri}}
\shortauthors{Johnson \& Gammie}
\shorttitle{Radially Stratified Disks}
\begin{document}

\title{Linear Theory of Thin, Radially-Stratified Disks}

\author{Bryan M. Johnson and Charles F. Gammie}

\affil{Center for Theoretical Astrophysics,
University of Illinois at Urbana-Champaign,
1110 West Green St., Urbana, IL 61801}

\begin{abstract}

We consider the nonaxisymmetric linear theory of radially-stratified
disks.  We work in a shearing-sheet-like approximation, where the
vertical structure of the disk is neglected, and develop equations for
the evolution of a plane-wave perturbation comoving with the shear flow
(a shearing wave, or ``shwave''). We calculate a complete solution set
for compressive and incompressive short-wavelength perturbations in both
the stratified and unstratified shearing-sheet models. We develop
expressions for the late-time asymptotic evolution of an individual
shwave as well as for the expectation value of the energy for an
ensemble of shwaves that are initially distributed isotropically in
$k$-space. We find that: (i) incompressive, short-wavelength
perturbations in the unstratified shearing sheet exhibit transient
growth and asymptotic decay, but the energy of an ensemble of such
shwaves is constant with time (consistent with \citealt {amn04}); (ii)
short-wavelength compressive shwaves grow asymptotically in the
unstratified shearing sheet, as does the energy of an ensemble of such
shwaves; (iii) incompressive shwaves in the stratified shearing sheet
have density and azimuthal velocity perturbations $\delta \Sigma$,
$\delta v_y \sim t^{-{\rm Ri}}$ (for $|{\rm Ri}| \ll 1$), where ${\rm
Ri} \equiv N_x^2/ (\qe \Omega)^2$ is the Richardson number, $N_x^2$ is
the square of the radial
Brunt-V$\ddot{\rm{a}}$is$\ddot{\rm{a}}$l$\ddot{\rm{a}}$ frequency and
$\qe \Omega$ is the effective shear rate; (iv) the energy of an ensemble
of incompressive shwaves in the stratified shearing sheet behaves
asymptotically as $\Ri \, t^{1-4{\rm Ri}}$ for $|{\rm Ri}| \ll 1$. For
Keplerian disks with modest radial gradients, $|{\rm Ri}|$ is expected
to be $\ll 1$, and there will therefore be weak growth in a single
shwave for ${\rm Ri} < 0$ and near-linear growth in the energy of an
ensemble of shwaves, independent of the sign of Ri.

\end{abstract}

\keywords{accretion, accretion disks, solar system: formation, galaxies:
nuclei}

\section{Introduction}

Angular momentum transport is central to the evolution of astrophysical
disks.  In many disks angular momentum is likely redistributed
internally by magnetohydrodynamic (MHD) turbulence driven by the
magnetorotational instability (MRI; see \citealt{bh98}).  But in portions of
disks around young, low-mass stars, in cataclysmic-variable disks in
quiescence, and in X-ray transients in quiescence \citep{sgbh00,gm98,men00}, 
disks may be composed of gas that is so neutral that the MRI fails.  It is 
therefore of interest to understand if there are purely hydrodynamic 
mechanisms for driving turbulence {\it and} angular momentum transport 
in disks.

The case for hydrodynamic angular momentum transport is not promising.
Numerical experiments carried out under conditions similar to those
under which the MRI produces ample angular momentum fluxes-- local
shearing-box models-- show small or negative angular momentum fluxes
when the magnetic field is turned off \citep{hgb95,hgb96}.  Unstratified 
shearing-sheet models show decaying angular momentum flux and kinetic 
energy when nonlinearly perturbed, yet recover the well known, high 
Reynolds number nonlinear instability of plane Couette flow when the 
parameters of the model are set appropriately (\citealt{bhs96}; see, however, 
the recent results by \citealt{ur04}). Local models 
with unstable vertical stratification show overturning and the development 
of convective turbulence, but the mean angular momentum flux is small and 
of the wrong sign \citep{sb96}.

Linear theory of global disk models has long indicated the presence of
instabilities associated with reflecting boundaries or features in the
flow (see e.g., \citealt{pp84,pp85,pp87,ggn86,gng87,ngg87,love99,li00}). 
Numerical simulations of the nonlinear outcome of these instabilities 
suggest that they saturate at low levels and are turned off by modest 
accretion \citep{bla87,haw91}.  One might guess that in the nonlinear 
outcome these instabilities will attempt to smooth out the features that 
give rise to them, much as convection tends to erase its parent entropy 
gradient.  There are some suggestions, however, that such instabilities saturate
into long-lived vortices, which may serve as obstructions in the flow that
give rise to angular momentum transport \citep{li01}.  We will consider
this possibility in a later publication.

Linear theory has yet to uncover a {\it local} instability of
hydrodynamic disks that produces astrophysically-relevant angular
momentum fluxes.  Because of the absence of a complete set of modes in
the shearing-sheet model, however, local linear stability is difficult
to prove.  Local nonlinear stability may be impossible to prove.
Comparison with laboratory Couette flow experiments is complicated by
several factors, not least of which is the inevitable presence of solid
radial boundaries in the laboratory that have no analogue in
astrophysical disks.

Recently, however, \cite{kb03} (hereafter KB03) have claimed to find a local 
hydrodynamic instability in global numerical simulations. The 
instability arises in a model with scale-free initial conditions (an 
equilibrium entropy profile that varies as a power-law in 
radius) and thus does not depend on sharp features in the flow. 
\cite{klr04} has performed a local linear stability analysis of a
radially-stratified accretion disk in an effort to explain the numerical
results obtained by KB03. The instability mechanism invoked is the
phenomenon of transient amplification as a shearing wave goes from
leading to trailing. This is the mechanism that operates for
nonaxisymmetric shearing waves in a disk that is nearly unstable to the
axisymmetric gravitational instability \citep{glb65,jt66,gt78}. It is the 
purpose of this work to clarify and extend the linear analysis of
\cite{klr04}. If this instability exists it could be important for the
evolution of low-ionization disks.

To isolate the cause for instabilities originally observed in global 3D
simulations, KB03 perform both local and global 2D calculations
in the ($R,\phi$)-plane. The local simulations use a new set of boundary
conditions termed the shearing-disk boundary conditions.  The model is
designed to simulate a local portion of the disk without neglecting
global effects such as curvature and horizontal flow gradients. The
boundary conditions, which are described in more detail in KB03,
require the assumption of a power-law scaling for the mean values of
each of the variables, as well as the assumption that the fluctuations
in each variable are proportional to their mean values. The radial
velocity component in the inner and outer four grid cells is damped by
$5\%$ each time step in order to remove artificial radial oscillations
produced by the model.\footnote{It is not surprising that shearing disk
boundary conditions as implemented in KB03 produce features on the
radial boundary, because the Coriolis parameter is discontinuous across
the radial boundary.}

The equilibrium profile for KB03's 2D runs was a constant surface density
$\Sigma$ with either a constant temperature $T$ or a temperature profile
$T \propto R^{-1}$. The constant-$T$ runs showed no instability while
those with varying $T$ (and thus varying entropy) sustained turbulence
and positive Reynolds stresses.\footnote{Notice that with a constant
$\Sigma$, the constant-$T$ runs have no variation in any of the
equilibrium variables, so it is not clear that the effects being
observed in the 2D calculations are due to the presence of an entropy
gradient rather than due simply to the presence of a pressure gradient.}
The fiducial local simulations were run at a resolution of $64^2$, with
a spatial domain of $R = 4$ to $6$ AU and $\Delta \phi = 30\,^{\circ}$.
The unstable run was repeated at a resolution of $128^2$, along with a
run at twice the physical size of the fiducial runs. One global model
(with non-reflecting outflow boundary conditions) was run at a
resolution of $128^2$ with a spatial domain of $R = 1$ to $10$ AU and
$\Delta \phi = 360\,^{\circ}$. All the runs yielded similar results,
with the larger simulations producing vortices and power on large
scales.

KB03 have chosen the term ``baroclinic instability'' by way of
analogy with the baroclinic instability that gives rise to weather
patterns in the atmosphere of the Earth and other planets (see e.g. 
\citealt{ped87}).\footnote{A baroclinic flow is one in which surfaces 
of constant density are inclined with respect to surfaces of constant 
pressure. If these surfaces coincide, the flow is termed barotropic.} 
The analogy is somewhat misleading, however, since the baroclinic 
instability that arises in planetary contexts is due to a baroclinic 
equilibrium. In a planetary atmosphere, a baroclinically-unstable 
situation requires stratification in both the vertical and latitudinal 
directions.\footnote{Contrary to the claim in \cite{klr04}, the 
two-layer model \citep{ped87} does not ignore the vertical structure; 
it simply considers the lowest-order vertical mode.} The stratification 
in KB03 is only in the radial direction, and as a result the
equilibrium is barotropic. It is the perturbations that are baroclinic; 
i.e., the disk is only baroclinic at linear order in the amplitude of a 
disturbance.

\cite{cab84} and \cite{ks86} have analyzed a thin disk with a baroclinic
equilibrium state (with both vertical and radial gradients). The latter
find that due to the dominant effect of the Keplerian shear, the
instability only occurs if the radial scale height is comparable to the
vertical scale height, a condition which is unlikely to be
astrophysically relevant.  As pointed out in KB03, the salient feature
that is common to their analysis and the classical baroclinic
instability is an equilibrium entropy gradient in the horizontal
direction. As we show in \S 2, however, an entropy gradient is not
required in order for two-dimensional perturbations to be baroclinic;
any horizontal stratification will do.

The ``global baroclinic instability'' claimed by KB03 is thus
analogous to the classical baroclinic instability in the sense that both
have the potential to give rise to convection.\footnote{The classical
baroclinic instability gives rise to a form of ``sloping convection''
\citep{hou02} since the latitudinal entropy gradient is inclined with
respect to the vertical buoyancy force.} When neglecting vertical
structure, however, the situation in an accretion disk is more closely
analogous to a shearing, stratified atmosphere, the stability of which
is governed by the classical Richardson criterion \citep{jwm61,chi70}. 
The only additional physics in a disk is the presence of the Coriolis force.
Most analyses of a shearing, stratified atmosphere, however, only 
consider stratification profiles that are stable to convection. The primary
question that \cite{klr04} and this work are addressing, then, is
whether or not the presence of shear stabilizes a stratified equilibrium 
that would be unstable in its absence.

We begin in \S 2 by outlining the basic equations for a local model of a
thin disk.  \S\S 3 and 4 describe the local linear theory for
nonaxisymmetric sinusoidal perturbations in unstratified and
radially-stratified disks, respectively.  We summarize and discuss the
implications of our findings in \S 5.

\section{Basic Equations}

The effect of radial gradients on the local stability of a thin disk can
be analyzed most simply in the two-dimensional shearing-sheet
approximation. This is obtained by a rigorous expansion of the equations
of motion in the ratio of the vertical scale height $H$ to the local radius $R$,
followed by a vertical integration of the fluid equations. The basic
equations that one obtains (e.g., \citealt{gt78}) are
\be\label{EQ1}
\dv{\Sigma}{t} + \Sigma \bnabla \cdot \bv = 0,
\ee
\be\label{EQ2}
\dv{\bv}{t} + {\bnabla P\over{\Sigma}} + 2\bO\times\bv - 2q\Omega^2 x \ex = 0,
\ee
\be\label{EQ3}
\dv{\,{\rm{ln}} S}{t} = 0,
\ee
where $\Sigma$ and $P$ are the two-dimensional density and pressure,
$S \equiv P \Sigma^{-\gamma}$ is monotonically related to the fluid
entropy,\footnote{With the assumptions of vertical hydrostatic equilibrium 
and negligible self-gravity, the effective two-dimensional adiabatic index 
can be shown to be $\gamma = (3\gamma_{3D} - 1)/(\gamma_{3D} + 1)$ 
(e.g. \citealt{ggn86}).} $\bv$ is the fluid velocity and $d/dt$ is the
Lagrangian derivative. The third and fourth terms in equation
(\ref{EQ2}) represent the Coriolis and centrifugal forces in the local
model expansion, where $\Omega$ is the local rotation frequency, $x$ is
the radial Cartesian coordinate and $q$ is the shear parameter (equal to
$1.5$ for a disk with a Keplerian rotation profile). The gravitational
potential of the central object is included as part of the centrifugal
force term in the local-model expansion, and we ignore the self-gravity
of the disk.

It is worth emphasizing at this point that we have integrated out the
vertical degrees of freedom in the model.  We will later focus on
perturbations with planar wavelengths that are small compared to a scale
height, and these perturbations will be strongly influenced by the
vertical structure of the disk.  

Equations (\ref{EQ1}) through (\ref{EQ3}) can be combined into a single
equation governing the evolution of the potential vorticity:
\be\label{PVEV}
\dv{}{t}\left(\frac{\bnabla \times \bv + 2\bO}{\Sigma}\right) \equiv
\dv{\bxi}{t} = \frac{\bnabla \Sigma \times \bnabla P}{\Sigma^3}.
\ee
In two dimensions, $\bxi$ has only one nonzero component and can
therefore be regarded as a scalar. Equation (\ref{PVEV}) demonstrates 
that for $P \equiv P(\Sigma)$ (as in the case of a strictly adiabatic evolution 
with isentropic initial conditions), the potential vorticity of fluid elements 
is conserved. For $P \neq P(\Sigma)$, however, the potential vorticity 
evolves with time. A barotropic equilibrium stratification can result in 
baroclinic perturbations that cause the potential vorticity to evolve at linear 
order. This can be seen by linearizing the scalar version of equation 
(\ref{PVEV}):
\be\label{PVEVLIN}
\pdv{\delta \xi}{t} + \bv_0 \cdot \bnabla \delta \xi + \delta \bv \cdot \bnabla \xi_0 = \frac{\ez \cdot(\bnabla 
\Sigma_0 \times \bnabla \delta P - \bnabla P_0 \times \bnabla \delta \Sigma)}{\Sigma_0^3},
\ee
where we have dropped the term $\propto \bnabla \Sigma_0 \times \bnabla
P_0$. Notice that an entropy gradient is not required for the evolution of 
the perturbed potential vorticity. For $S_0 = P_0 \Sigma_0^\gamma = 
constant$, equation (\ref{PVEVLIN}) reduces to
\be\label{PVEVLIN2}
\pdv{\delta \xi}{t} + \bv_0 \cdot \bnabla \delta \xi + \delta \bv \cdot \bnabla \xi_0 = \frac{\ez \cdot(\bnabla 
P_0 \times \bnabla \delta S)}{\gamma \Sigma_0^2 S_0}.
\ee
Potential vorticity is conserved only in the limit of zero stratification ($P_0 
= constant$) or adiabatic perturbations ($\delta S = 0$).

\section{Unstratified Shearing Sheet}

Our goal is to understand the effects of radial stratification, but we
begin by developing the linear theory of the standard (unstratified)
shearing sheet, in which the equilibrium density and pressure are
assumed to be spatially constant.  This will serve to establish notation
and method of analysis and to highlight the changes introduced by
radial stratification in the next section.

Our analysis follows that of \cite{gt78} except for our neglect of
self-gravity. The equilibrium consists of a uniform sheet with $\Sigma =
\Sigma_0 = constant$, $P = P_0 = constant$, and $\bv_0 = -q\Omega x
\ey$. We consider nonaxisymmetric Eulerian perturbations about this
equilibrium with space-time dependence $\delta(t){\rm exp} (ik_x(t) x +
ik_y y)$, where
\be\label{KX}
k_x(t) \equiv k_{x0} + q\Omega k_y t
\ee
(with $k_{x0}$ and $k_y > 0$ constant) is required to allow for a spatial Fourier 
decomposition of the perturbation. We will refer to these perturbations 
as shearing waves, or with some trepidation, but more compactly, as 
{\it shwaves}.

\subsection{Linearized Equations}

To linear order in the perturbation amplitudes, the dynamical equations
reduce to
\be\label{LIN1s}
\frac{\dot{\delta \Sigma}}{\Sigma_0} + i k_x \delta v_x 
	+ i k_y \delta v_y = 0,
\ee
\be\label{LIN2s}
\dot{\delta v}_x - 2\Omega \delta v_y 
	+ ik_x \frac{\delta P}{\Sigma_0} = 0,
\ee
\be\label{LIN3s}
\dot{\delta v}_y + (2- q)\Omega \delta v_x 
	+ ik_y \frac{\delta P}{\Sigma_0} = 0,
\ee
\be\label{LIN4s}
\frac{\dot{\delta P}}{\Sigma_0} + c_s^2 (i k_x \delta v_x 
	+ i k_y \delta v_y)  = 0,
\ee
where $c_s^2 = \gamma P_0/\Sigma_0$ is the square of the equilibrium
sound speed and an over-dot denotes a time derivative.

The above system of equations admits four linearly-independent
solutions. Two of these are the non-vortical shwaves (solutions for 
which the perturbed potential vorticity is zero), which in the absence 
of self-gravity can be solved for exactly. The remaining two solutions are 
the vortical shwaves. When $k_y \rightarrow 0$ the latter reduce to the 
zero-frequency modes of the axisymmetric version of equations 
(\ref{LIN1s}) through (\ref{LIN4s}). One of these (the entropy mode) 
remains unchanged in nonaxisymmetry (in a frame comoving with the 
shear). There is thus only one nontrivial vortical shwave in the 
unstratified shearing sheet.

In the limit of tightly-wound shwaves ($|k_x| \gg k_y$), the nonvortical 
and vortical shwaves are compressive and incompressive, respectively. 
In the short-wavelength limit ($H k_y \gg 1$, where $H \equiv c_s/\Omega$ 
is the vertical scale height), the compressive and incompressive solutions  
remain well separated at all times, but for $H k_y \lesssim O(1)$ there is mixing 
between them near $k_x = 0$ as an incompressive shwave shears from 
leading to trailing.\footnote{This was pointed out to us by J. Goodman.} 
With the understanding that the distinction between compressive shwaves 
and incompressive shwaves as separate solutions is not valid for all time 
when $H k_y \lesssim O(1)$, we generally choose to employ these terms over 
the more general but less intuitive terms ``non-vortical'' and ``vortical.''

Based upon the above considerations, it is convenient to study the 
vortical shwave in the short-wavelength, low-frequency ($\partial_t \ll 
c_s k_y$) limit. This is equivalent to working in the Boussinesq 
approximation,\footnote{We demonstrate this equivalence in the Appendix.} 
which in the unstratified shearing sheet amounts to assuming incompressible 
flow. In this limit, equation (\ref{LIN1s}) is replaced with
\be\label{LIN1b}
k_x \delta v_x + k_y \delta v_y = 0.
\ee
This demonstrates the incompressive nature of the vortical shwave in 
the short-wavelength limit.

\subsection{Solutions}

In the unstratified shearing sheet, equation (\ref{PVEVLIN}) for the 
perturbed potential vorticity can be integrated to give:
\be\label{POTVORT}
\delta \xi_u = \frac{i k_x \delta v_y - i k_y \delta v_x}{\Sigma_0} - 
\xi_0 \frac{\delta \Sigma}{\Sigma_0} = constant,
\ee
where $\xi_0 = (2 - q)\Omega/\Sigma_0$ is the equilibrium potential
vorticity and we have employed the subscript $u$ to highlight the fact 
that the perturbed potential vorticity is only constant in the unstratified 
shearing sheet. To obtain the compressive-shwave solutions, we set the 
constant $\delta \xi_u$ to zero. Combining equations (\ref{LIN3s}) and
(\ref{POTVORT}) with $\delta \xi_u = 0$, one obtains an expression for
$\delta v_{xc}$ in terms of $\delta v_{yc}$ and its derivative:
\be\label{CVX}
\delta v_{xc} = \frac{c_s^2 k_x k_y \delta v_{yc} - \xi_0 \Sigma_0
\dot{\delta v}_{yc}}{\xi_0^2 \Sigma_0^2 + c_s^2 k_y^2},
\ee
where the subscript $c$ indicates a compressive shwave.  The associated 
density and pressure perturbations are
\be\label{CS}
\delta \Sigma_c = \frac{\delta P_c}{c_s^2} = i \frac{\xi_0 \Sigma_0 k_x 
\delta v_{yc} + k_y \dot{\delta v}_{yc}}{\xi_0^2 \Sigma_0^2 + c_s^2 k_y^2}
\ee
via equation (\ref{POTVORT}). Reinserting equation (\ref{CVX}) into
equation (\ref{LIN3s}), taking one time derivative and replacing
$\dot{\delta P}$ via equation (\ref{LIN4s}), we obtain the following
remarkably simple equation:
\be\label{SOUND}
\ddot{\delta v}_{yc} + \left(c_s^2 k^2 + \kappa^2\right) \delta v_{yc} = 0,
\ee
where $k^2 = k_x^2 + k_y^2$ and $\kappa^2 = (2 - q)\Omega^2$ is the
epicyclic frequency. Changing to the dimensionless dependent variable
\be\label{TTAU}
T \equiv i \sqrt{\frac{2 c_s k_y}{q\Omega}} \left(q \Omega t + \frac{k_{x0}}
{k_y}\right) \equiv i \sqrt{\frac{2 c_s k_y}{q\Omega}} \, \tau
\ee
and defining
\be
C \equiv \frac{c_s^2 k_y^2 + \kappa^2}{2 q\Omega c_s k_y},
\ee
the equation governing $\delta v_{yc}$ becomes
\be\label{VYT}
\dv{^2\delta v_{yc}}{T^2} + \left(\frac{1}{4}T^2 - C\right) \delta v_{yc} = 0.
\ee
This is the parabolic cylinder equation (e.g. \citealt{as72}), 
the solutions of which are parabolic cylinder functions. One 
representation of the general solution is
\be\label{CVY}
\delta v_{yc} = e^{-\frac{i}{2}T^2}\left[c_1 \, M\left(\frac{1}{4} - 
\frac{i}{2}C,\frac{1}{2},\frac{i}{2}T^2\right) + c_2 \, T \, M\left(
\frac{3}{4} - \frac{i}{2}C,\frac{3}{2},\frac{i}{2}T^2\right)\right],
\ee
where $c_1$ and $c_2$ are constants of integration and $M$ is a 
confluent hypergeometric function. This completely specifies the 
compressive solutions for the unstratified shearing sheet, for any 
value of $k_y$.

Equation (\ref{VYT}) has been analyzed in detail by \cite{ngg87}; 
their modal analysis yields the analogue of equation (\ref{VYT}) in 
radial-position space rather than in the radial-wavenumber ($k_x = 
k_y \tau$) space that forms the natural basis for our shwave 
analysis. One way of seeing the correspondence between the 
modes and shwaves is to take the Fourier transform of the asymptotic 
form of the solution. Appropriate linear combinations of the solutions 
given in equation (\ref{CVY}) have the following asymptotic time 
dependence for $\tau \gg 1$:
\be
\delta v_{yc} \propto \sqrt{\frac{2}{T}}\exp\left(\pm \frac{i}{4}T^2 
\right) \propto \frac{1}{\sqrt{k_x}}\exp\left(\pm i \int c_s k_x 
\, dt \right).
\ee
The Fourier transform of the above expression, evaluated by the method of 
stationary phase for $H k_y \gg 1$, yields
\be
\delta v_{yc}(X) \propto \sqrt{\frac{2}{X}}\exp\left(\pm \frac{i}{4}X^2\right),
\ee
which is equivalent to the expressions given for the modes analyzed by 
\cite{ngg87}, in which the dimensionless spatial variable (with zero frequency, 
so that corotation is at $x = 0$) is defined as
\be
X \equiv \sqrt{\frac{2 q \Omega k_y}{c_s}} x.
\ee

To obtain the incompressive shwave, we use the condition of incompressibility
(equation (\ref{LIN1b})) to write $\delta v_y$ in terms of $\delta v_x$,
and then combine the dynamical equations (\ref{LIN2s}) and (\ref{LIN3s})
to eliminate $\delta P$. The incompressive shwave is given by:
\be\label{IVX}
\delta v_{xi} = \delta v_{xi0}\frac{k_0^2}{k^2},
\ee
\be\label{IVY}
\delta v_{yi} = -\frac{k_x}{k_y} \delta v_{xi},
\ee
\be\label{IS}
\frac{\delta \Sigma_i}{\Sigma_0} = \frac{\delta P_i}{\gamma P_0} = 
\frac{1}{i c_s k_y}\left(\frac{k_x}{k_y} \frac{\dot{\delta v}_{xi}}{c_s} 
+ 2(q - 1) \Omega \frac{\delta v_{xi}}{c_s}\right),
\ee
where the subscript $i$ indicates an incompressive shwave, $k_0^2 =
k_{x0}^2 + k_y^2$ and $\delta v_{xi0}$ is the value of $\delta v_{xi}$
at $t=0$.\footnote{\cite{cz03} obtained this solution by starting with
the assumption of incompressibility.} This solution is uniformly valid  
for all time to leading order in $(H k_y)^{-1} \ll 1$.

\subsection{Energetics of the Incompressive Shwaves}

We define the kinetic energy in a single incompressive shwave as
\be
E_{ki} \equiv \frac{1}{2}\Sigma_0 (\delta v_{xi}^2 + \delta v_{yi}^2) =
\frac{1}{2}\Sigma_0 \delta v_{xi}^2 \frac{k^2}{k_y^2} =
\frac{1}{2}\Sigma_0 \delta v_{xi0}^2 \frac{k_0^4}{k_y^2 k^2},
\ee
which peaks at $k_x = 0$. This is not the only possible 
definition for the energy associated with a shear-flow disturbance; 
see the discussion in Appendix A of \cite{ngg87}.

One can also define an amplification factor for an individual shwave,
\be
{\cal A } \equiv \frac{E_{ki}(k_x = 0)}{E_{ki}(t = 0)} =
1 + \frac{k_{x0}^2}{k_y^2},
\ee
which indicates that an arbitrary amount of transient amplification in 
kinetic energy can be obtained as one increases the amount of swing for 
a leading shwave ($k_{x0} \ll -k_y$). This is essentially the mechanism 
invoked by \cite{cz03}, \cite{ur04} and \cite{amn04} to argue for the onset 
of turbulence in unmagnetized Keplerian disks.

Because only a small subset of all Fourier components achieve large
amplification (those with initial wavevector very nearly aligned with
the radius vector), one must ask what amplification is achieved for an
astrophysically relevant set of initial conditions containing a
superposition of Fourier components.  It is natural to draw such a set
of Fourier components from a distribution that is isotropic, or nearly
so, when $k_0$ is large.

Consider, then, perturbing a disk with a random set of incompressive
perturbations (initial velocities perpendicular to $\bk_0$) drawn from an
isotropic, Gaussian random field and asking how the expectation value for the
kinetic energy associated with the perturbations evolves with time.  The
evolution of the expected energy density is given by the following
integral:
\be
\<E_i\> = L^2 \int d^2k_0 \<E_{ki}\> 
= L^2 \int d^2k_0 \frac{1}{2}\Sigma_0
\< \delta v_{xi0}^2\> \frac{k_0^4}{k_y^2 k^2}.
\ee
where $\<\>$ indicates an average over an ensemble of initial
conditions, the first equality follows from Parseval's theorem, the
second equality follows from the incompressive shwave solution
(\ref{IVX})-(\ref{IS}) and therefore applies only for $k_0 H \gg 1$,
and $L^2$ is a normalizing factor with units of length squared.

For initial conditions that are isotropic in $\bk_0$ ($\delta v_{xi0} =
\delta v_\perp (k_0,\theta) \sin \theta$, where $\<\delta v_\perp^2
(k_0)\>$ is the expectation value for the initial incompressive
perturbation as a function of $k_0$ and $\tan\theta=k_y/k_{x0}$), the
integral becomes
\be\label{EKI}
\< E_i\> = \frac{1}{2}\Sigma_0 L^2 \int k_0 dk_0 \< \delta v_\perp^2 (k_0)\>
\int_0^{2\pi} d\theta \, \frac{1}{\sin^2\theta + (q\Omega t \sin\theta 
+ \cos\theta)^2}.
\ee
Changing integration variables to $\tau = q\Omega t + \cot\theta$, the angular 
integral becomes
\be
\int_{-\infty}^{\infty} d\tau \, \frac{2}{1 + \tau^2} = 2\pi,
\ee
which is independent of time; hence
\be
\< E_i\> = \< E_i (t = 0)\>
\ee
and we do not expect the total energy in incompressive shwaves to evolve.

Although this result may appear to depend in detail on the assumption of
isotropy, one can show that it really only depends on $\< E_{ki} (t =
0)\>$ being smooth near $\sin \theta = 0$, i.e. that there should not be
a concentration of power in nearly radial wavevectors. This can be seen
from the following argument. If we relax the assumption of isotropy, the
angular integral becomes
\be
\int_0^{2\pi} d\theta \, \frac{\< \delta v_\perp^2(k_0,\theta) \>}
{\sin^2\theta + (q\Omega t \sin\theta + \cos\theta)^2}.
\ee
For $q\Omega t \gg 1$ the above integrand is sharply peaked in the
narrow regions around $\tan \theta = -1/(q\Omega t) \ll 1$ (i.e., $\sin
\theta \simeq 0$). One can perform a Taylor-series expansion of $\<
\delta v_\perp^2 (k_0,\theta) \>$ in these regions, and as long as $\<
\delta v_\perp^2(k_0,\theta) \>$ itself is not sharply peaked it is well
approximated as a constant. A modest relaxation of the assumption of
isotropy, then, will result in an asymptotically constant value for the 
energy integral.

Based upon this analysis, large amplification in an individual shwave
does not in itself argue for a transition to turbulence due to transient
growth.  One must also demonstrate that a ``natural'' set of
perturbations can extract energy from the background shear flow.  In the
case of the unstratified shearing sheet, the energy of a random set of
incompressive perturbations remains constant with time.  This is
consistent with the results of \cite{ur04}, who see asymptotic decay in
linear theory, because they work with a finite set of wavevectors, each
of which must decay asymptotically.

\subsection{Energetics of the Compressive Shwaves}

Here we calculate the energy evolution of the compressive shwaves
for comparison purposes. We will consider the evolution of 
short-wavelength compressive shwaves in which only the initial velocity is 
perturbed, both for simplicity and for consistency with our calculation 
of the short-wavelength incompressive shwaves.  As before, we
will assume that the initial kinetic energy is distributed isotropically.

We use the WKB solutions to equation (\ref{SOUND}) with $H k_y \gg 
1$.\footnote{These solutions are the short-wavelength, {\it high}-frequency 
($\partial_t \sim O(c_s k_y)$) limit of the full set of linear equations in 
the shearing sheet; see the Appendix.}  With the initial density perturbation 
set to zero (consistent with our assumption of only initial velocity 
perturbations), the uniformly-valid asymptotic solution to leading order 
in $(H k_y)^{-1}$ is given by
\be\label{VYWKB}
\delta v_{yc} = \delta v_{yc0} \sqrt{\frac{k_0}{k}} \cos(W-W_0),
\ee
\be\label{VXWKB}
\delta v_{xc} = \frac{k_x}{k_y} \delta v_{yc},
\ee
\be\label{SWKB}
\delta \Sigma_c = \frac{i}{c_s^2 k_y}\dot{\delta v}_{yc},
\ee
where the WKB eikonal is given by
\be
W \equiv \int c_s k \, dt = \frac{H k_y}{q} \int \sqrt{1 + \tau^2} \, 
d\tau = \frac{H k_y}{2 q} \left(\tau \sqrt{1 + \tau^2} + \ln\left(\tau + 
\sqrt{1 + \tau^2}\right) \right),
\ee
with $W_0$ being the value of $W$ at $t=0$.\footnote{This is
not the same WKB solution that is calculated in the tight-winding
approximation by \cite{gt78}; in that case $c_s k_y/\kappa \ll 1$, the
opposite limit to that which we are considering here. The two WKB 
solutions match for $\tau \gg 1$ in the absence of self-gravity. We have 
verified the accuracy of this solution by comparing it to the exact 
solution with acceptable results, and it is valid to leading order for all time.}

Using equation (\ref{VXWKB}), the energy integral for the compressive 
shwaves in the short-wavelength limit is
\be
\< E_{c}\> = L^2 \int d^2k_0 \<E_{kc}\> = L^2  
\int d^2k_0 \frac{1}{2}\Sigma_0 \< \delta v_{yc}^2 \>\frac{k^2}{k_y^2}.
\ee
With initial velocities now parallel to ${\bk}_0$ (and again isotropic), 
this becomes
\be
\< E_{c}\> = 
\frac{1}{2}\Sigma_0 L^2
\int k_0 dk_0 \< \delta v_\|^2(k_0) \>
\int_0^{2\pi} d\theta \,
\sqrt{\sin^2\theta + (q\Omega t \sin\theta + \cos\theta)^2}\cos^2(W-W_0).
\ee
For $q \Omega t \gg 1$, the angular integral is approximated by
\be
\int_0^{2\pi} d\theta \, |\sin\theta| \left(1 + \cos(2W-2W_0)\right) \simeq 
2 q \Omega t + \sqrt{\frac{2\pi q \Omega}{c_s k_0}} \cos(c_s k_0 q 
\Omega t^2 - \pi/4),
\ee
where the second approximation comes from employing the method of
stationary phase.\footnote{The first approximation breaks down near
$\sin \theta = 0$, but the contribution of these regions to the integral
is negligible for $q \Omega t \gg 1$, in contrast to the situation for
incompressive shwaves.} In the short-wavelength limit, then,
\be
\< E_{c} (q \Omega t \gg 1) \> =  2 q \Omega t \, \< E_{c}(t = 0) \>.
\ee
Thus the kinetic energy of an initially isotropic distribution of compressive
shwaves grows, presumably at the expense of the background shear flow.  

The fate of a single compressive shwave is to steepen into a weak shock
train and then decay.  The fate of the field of weak shocks generated by
an ensemble of compressive shwaves is less clear, but the mere presence
of weak shocks does not indicate a transition to turbulence.

\section{Radially-Stratified Shearing Sheet}

We now generalize our analysis to include the possibility that the
background density and pressure varies with $x$; this stratification is
required for the manifestation of a convective instability.  In order to
use the shwave formalism we must assume that the background
varies on a scale $L \sim H \ll R$ so that the local model expansion
(e.g., the neglect of curvature terms in the equations of motion) is
still valid.

With this assumption the equilibrium condition becomes
\be\label{V0}
\bv_0 = \left(-q\Omega x + \frac{P_0^\prime(x)}{2\Omega 
\Sigma_0(x)}\right)\ey,
\ee
where a prime denotes an $x$-derivative. 
One can regard the background flow as providing an effective shear rate
\be
\qe \Omega \equiv -v_0^\prime
\ee
that varies with $x$, in which case $\bv_0 = -\int^x \qe(s) ds \, \Omega
\ey$. 

Localized on this background flow we will consider a shearing
wave with $k_y L \gg 1$.  That is, we will consider nonaxisymmetric
short-wavelength Eulerian perturbations with spacetime dependence
$\delta(t) \exp(i\int^x \tilde{k}_x(t,s) ds + ik_y y + ik_z z)$,
where $k_y$ and $k_z$ are constants and
\be
\tilde{k}_x(t,x) \equiv k_{x0} + \qe(x)\Omega k_y t.
\ee  

It may not be immediately obvious that this is a valid expansion since 
the shwaves sit on top of a radially-varying background (see \citealt{toom69} 
for a discussion of waves in a slowly-varying background).  But
this is an ordinary WKB expansion in disguise.  To see this, one
need only transform to ``comoving'' coordinates $x' = x$, $y'
= y + \int^x \qe(s) ds \Omega t$, $t' = t$ (this procedure may be more familiar
in a cosmological context; as \cite{bal88} has pointed out, this
is possible for any flow in which the velocities depend linearly
on the spatial coordinates).  In this frame the time-dependent
wavevector given above is transformed to a time-independent wavevector.
The price paid for this is that $\partial_x \rightarrow
\partial_{x'} + \qe \Omega t \partial_{y'}$, so new explicit time
dependences appear on the right hand side of the perturbed equations
of motion, and the perturbed variables no longer have time dependence
$\exp(i\omega t')$.  Instead, we must solve an ODE for $\delta(t')$.
The $y'$ dependence can be decomposed as $\exp(ik_y y')$. The 
$x'$ dependence can be treated via WKB, since the perturbation may 
be assumed to have the form $W(\epsilon x', \epsilon t')\, \exp(i\bk' 
\cdot \bx')$. This ``nearly diagonalizes'' the operator $\partial_{x'}$.
Thus we are considering the evolution of a wavepacket in comoving
coordinates--- a ``shwavepacket''.

For this procedure to be valid two conditions must be met.  First the
usual WKB condition must apply, $k_y L \gg 1$.  Second, the parameters
of the flow that are ``seen'' by the shwavepacket must change little on
the characteristic timescale for variation of $\delta(t)$, which is
$\Omega^{-1}$ for the incompressive shwaves.  For solid body rotation
($\tilde{q} = 0$) the group velocity (derivable from equation (\ref{DRQ0}),
below) is $|v_g| < N_x/k$ (for positive squared Brunt-V\"ais\"al\"a
frequency $N_x^2$, defined below; for $N_x^2 < 0$ the waves grow in place),
so the timescale for change of wave packet parameters in this case is
$L/|v_g| > k L/N_x \gg \Omega^{-1}$.  It seems reasonable to anticipate
similarly long timescales when shear is present.  As a final check, we
have verified directly, using a code based on the ZEUS code of
\cite{sn92}, that a vortical shwavepacket in the stratified shearing
sheet remains localized as it swings from leading to trailing.

\subsection{Linearized Equations}

To linear order in the perturbation amplitudes, the dynamical equations
reduce to
\be\label{LIN1a}
\frac{\dot{\delta \Sigma}}{\Sigma_0} + \frac{\delta v_x}{L_{\Sigma}}  
+ i \tilde{k}_x \delta v_x + i k_y \delta v_y + i k_z \delta v_z = 0,
\ee
\be\label{LIN2}
\dot{\delta v}_x - 2\Omega \delta v_y + i\tilde{k}_x \frac{\delta P}{\Sigma_0}  -
\frac{c_s^2}{L_P} \frac{\delta \Sigma}{\Sigma_0} = 0,
\ee
\be\label{LIN3}
\dot{\delta v}_y + (2- \qe)\Omega \delta v_x + ik_y \frac{\delta P}{\Sigma_0} = 0,
\ee
\be\label{LIN4}
\dot{\delta v}_z + ik_z \frac{\delta P}{\Sigma_0} = 0,
\ee
\be\label{LIN5a}
\frac{\dot{\delta P}}{\Sigma_0} - c_s^2 \frac{\dot{\delta \Sigma}}
{\Sigma_0} + c_s^2 \frac{\delta v_x}{L_S} = 0,
\ee
where
\be
\frac{1}{L_P} \equiv \frac{P_0^\prime}{\gamma P_0} = 
\frac{1}{L_{\Sigma}} + \frac{1}{L_S} \equiv
\frac{\Sigma_0^\prime}{\Sigma_0} +
\frac{S_0^\prime}{\gamma S_0}
\ee
define the equilibrium pressure, density and entropy scale heights.
We have included the vertical component of the velocity in
order to make contact with an axisymmetric convective instability that
is present in two dimensions, after which we will set $k_z$ to zero. 

We will be mainly interested in the incompressive shwaves because the
short-wavelength compressive shwaves are unchanged at leading order by
stratification. We will therefore work solely in the Boussinesq
approximation.\footnote{We also drop the subscripts distinguishing
between the compressive and incompressive shwaves.} In addition to the
assumption of incompressibility, this approximation considers $\delta P$
to be negligible in the entropy equation; pressure changes are
determined by whatever is required to maintain nearly incompressible
flow.  The original Boussinesq approximation applies only to
incompressible fluids. It was extended to compressible fluids by
\cite{jeff30} and \cite{sv60}. We show in the Appendix that it is
formally equivalent to taking the short-wavelength, low-frequency limit
of the full set of linear equations. From this viewpoint, assuming that
$H k_y \delta P/P_0$ is of the same order as the other terms in the
dynamical equations implies that $\delta P/P_0 \sim (H k_y)^{-1} \delta
\Sigma/\Sigma_0$, thus justifying its neglect in the entropy equation.
We therefore replace equations (\ref{LIN1a}) and (\ref{LIN5a}) with
\be\label{LIN1}
\tilde{k}_x \delta v_x + k_y \delta v_y + k_z \delta v_z = 0
\ee
and
\be\label{LIN5}
\frac{\dot{\delta \Sigma}}{\Sigma_0} - \frac{\delta v_x}{L_S} = 0.
\ee

Using equations (\ref{LIN3}) and (\ref{LIN5}) and the time derivative of
equation (\ref{LIN1}), one can express $\dot{\delta v}_y$ and $\delta P$
in terms of $\delta v_x$ and $\dot{\delta v}_x$:
\be\label{DH}
\frac{\delta P}{\Sigma_0} = -i \frac{\tilde{k}_x \dot{\delta v}_x + 2(\qe - 1) 
\Omega k_y \delta v_x}{k_y^2 + k_z^2},
\ee
\be\label{DVY}
\dot{\delta v}_y = \frac{(-\qe k_y^2 + (\qe - 2)k_z^2) \Omega \delta v_x - 
\tilde{k}_x k_y \dot{\delta v}_x}{k_y^2 + k_z^2}.
\ee
Eliminating $\delta P$ in equation (\ref{LIN2}) via equation (\ref{DH}) gives
\be
\tilde{k}^2 \dot{\delta v}_x + 2(\qe - 1) \Omega \tilde{k}_x k_y \delta v_x = (k_y^2 + k_z^2) (2 \Omega \delta v_y + (c_s^2/L_P) \delta \Sigma/\Sigma_0),
\ee
where $\tilde{k}^2 = \tilde{k}_x^2 + k_y^2 + k_z^2$. Taking the time
derivative of this equation and eliminating $\dot{\delta \Sigma}$ and
$\dot{\delta v}_y$ via equations (\ref{LIN5}) and (\ref{DVY}), we obtain
the following differential equation for $\delta v_x$:
\be\label{BOUSSVX}
\tilde{k}^2 \ddot{\delta v}_x + 4 \qe \Omega \tilde{k}_x k_y \dot{\delta v}_x 
+ \left[k_y^2\left(N_x^2 + 2\qe^2 \Omega^2\right) + k_z^2\left(N_x^2 +
 \tilde{\kappa}^2\right)\right]\delta v_x = 0,
\ee
where $\tilde{\kappa}^2 = 2(2-\qe)\Omega^2$ is the square of the
effective epicyclic frequency and 
\be
N_x^2 \equiv -{c_s^2\over{L_S L_P}}
\ee
is the square of the
Brunt-V$\ddot{\rm{a}}$is$\ddot{\rm{a}}$l$\ddot{\rm{a}}$ frequency in the
radial direction.\footnote{Notice that $N_x^2$, $\qe$ and
$\tilde{\kappa}^2$ are all functions of $x$ and vary on a scale $L \sim
H$.}

\subsection{Comparison with Known Results}

Setting $k_y = 0$ in equation (\ref{BOUSSVX}) yields the axisymmetric
modes with the following dispersion relation (for $\delta(t) \propto
e^{-i\omega t}$):
\be
\omega^2 = \frac{k_z^2}{k_{x0}^2 + k_z^2}\left(N_x^2 + \tilde{\kappa}^2\right).
\ee
This is the origin of the H{\o}iland stability criterion: the axisymmetric
modes are stable for $N_x^2 + \tilde{\kappa}^2 > 0$. In the absence of
rotation this reduces to the Schwarzschild stability criterion: $N_x^2 >
0$ is the necessary condition for stability. The effect of rotation is strongly 
stabilizing: if $N_x^2 < -\tilde{\kappa}^2$, as required for instability, then 
$L_S L_P \sim H^2$; pressure and entropy must vary on radial scales of 
order the scale height for the disk to be H{\o}iland unstable.

Notice that effective epicyclic frequency $\tilde{\kappa}^2$ only stabilizes 
modes with nonzero $k_z$. The stability of nonaxisymmetric
shwaves with $k_z = 0$ (as in the mid-plane of a thin disk) is the open 
question that this work is addressing. In this limit and in
the absence of shear the Schwarzschild stability criterion is again
recovered: with $k_z = 0$ and $\qe = 0$ in equation (\ref{BOUSSVX}) the
dispersion relation becomes
\be\label{DRQ0}
\omega^2 = \frac{k_y^2}{k_{x0}^2 + k_y^2}N_x^2,
\ee
If there is a region of the disk where the effective shear is zero, a
WKB normal-mode analysis will yield the above dispersion relation and
there will be convective instability for $N_x^2 < 0$. It appears from
equation (\ref{BOUSSVX}) that differential rotation provides a
stabilizing influence for nonaxisymmetric shwaves just as rotation
does for the axisymmetric modes. Things are not as simple in
nonaxisymmetry, however. The time dependence is no longer exponential,
nor is it the same for all the perturbation variables. There is no clear
cutoff between exponential and oscillatory behavior, so the question of
flow stability becomes more subtle. 

As discussed in the introduction, the Boussinesq system of equations in 
the shearing-sheet model of a radially-stratified disk bear a close resemblance 
to the system of equations employed in analyses of a shearing, stratified 
atmosphere. A sufficient condition for stability in the latter case is that 
\be\label{RICH}
{\rm Ri} \equiv \frac{N_x^2}{(v_0^\prime)^2} \geq \frac{1}{4}
\ee
everywhere in the flow, where Ri is the Richardson number, a measure of 
the relative importance of buoyancy and shear. This stability criterion was 
originally proved by \cite{jwm61} and \cite{how61} for incompressible 
fluids, and its extension to compressible fluids was demonstrated by 
\cite{chi70}. The stability criterion is based on a normal-mode analysis 
with rigid boundary conditions. Other than differences in notation (e.g., 
our radial coordinate corresponds to the vertical coordinate in a stratified 
atmosphere), the key differences in our system are: (i) the equilibrium 
pressure gradient in a disk is balanced by centrifugal forces rather than 
by gravity; (ii) the disk equations contain Coriolis force terms; (iii) most 
atmospheric analyses only consider an equilibrium that is convectively 
stable, whereas we are interested in an unstable stratification; (iv) we
do not employ boundary conditions in our analytic model since we are 
only interested in the possibility of a local instability. 

The lack of boundary conditions in our model makes the applicability of 
the standard Richardson stability criterion in determining local stability 
somewhat dubious, since the lack of boundary conditions precludes 
the decomposition of linear disturbances into normal modes. 
The natural procedure for performing a local linear analysis in disks 
is to decompose the perturbations into shwaves, as we have done.

\cite{ehr53} consider both stable and unstable atmospheres and analyze 
an initial-value problem by decomposing the perturbations in time via
Laplace transforms.\footnote{Cited in \cite{jwm61}.} For flow between 
two parallel walls, they find that an arbitrary initial disturbance behaves 
asymptotically as $t^{(\alpha - 1)/2}$ for $-3/4 < {\rm Ri} < 1/4$, where
\be\label{ALPHA}
\alpha \equiv \sqrt{1 - 4 \, {\rm{Ri}}},
\ee
which grows algebraically for ${\rm Ri} < 0$. The disturbance grows 
exponentially only for ${\rm Ri} < -3/4$. For a semi-infinite flow, the 
power-law behavior in time holds for $-2 < {\rm Ri} < 1/4$, with 
exponential growth for ${\rm Ri} < -2$. These results illustrate the 
importance of boundary conditions in determining stability.

In the $k_z = 0$ limit that we are concerned with here, the correspondence 
between the disk and atmospheric models turns out to be exact in the 
shwave formalism. This is because the Coriolis force only appears in 
equation (\ref{BOUSSVX}) via $\tilde{\kappa}^2$, which disappears 
when $k_z = 0$. The equation describing the time evolution of shwaves 
in both a radially-stratified disk and a shearing, stratified atmosphere is thus
\be\label{BOUSSVX2D}
\tilde{k}^2 \ddot{\delta v}_x + 4 \qe \Omega \tilde{k}_x k_y \dot{\delta v}_x 
+ k_y^2\left(N_x^2 + 2\qe^2 \Omega^2\right)\delta v_x = 0.
\ee
We analyze the solutions to this equation in the following section.

\subsection{Solutions}

Changing time variables in equation (\ref{BOUSSVX2D}) to 
$\tilde{\tau} \equiv \tilde{k}_x/k_y$, the
differential equation governing $\delta v_x$ becomes
\be\label{ODE}
(1 + \tilde{\tau}^2)\dv{^2\delta v_x}{\tilde{\tau}^2} + 4 \tilde{\tau} 
\dv{\delta v_x}{\tilde{\tau}} + ({\rm{Ri}} + 2)\delta v_x = 0.
\ee
The solutions to equation (\ref{ODE}) are
hypergeometric functions. With the change of variables $z \equiv
-\tilde{\tau}^2$, equation (\ref{ODE}) becomes
\be\label{ODEZ}
z(1-z)\dv{^2\delta v_x}{z^2} + \frac{1- 5z}{2}\dv{\delta v_x}{z} - 
\frac{{\rm{Ri}} + 2}{4}\delta v_x = 0.
\ee
The hypergeometric equation \citep{as72}
\be\label{HGEQ}
z(1-z)\dv{^2\delta v_x}{z^2} + \left[c - (a + b + 1)z\right]\dv{\delta v_x}{z} 
- ab\delta v_x = 0
\ee
has as its two linearly independent solutions $F(a,b;c;z)$ and $z^{1-c}
F(a-c+1,b-c+1;2-c;z)$. Comparison of equations (\ref{ODEZ}) and 
(\ref{HGEQ}) shows that $a = (3 - \alpha)/4$, $b = (3 + \alpha)/4$ and 
$c = 1/2$, where $\alpha$ is defined in equation (\ref{ALPHA}).

The general solution for $\delta v_x$ is thus given by
\be\label{SOLVX}
\delta v_x = C_1 \,F\left(\frac{3 - \alpha}{4},\frac{3 + \alpha}{4};
\frac{1}{2};-\tilde{\tau}^2\right) + C_2 \, \tilde{\tau}\,F\left(\frac{
5 - \alpha}{4},\frac{5 + \alpha}{4};\frac{3}{2};-\tilde{\tau}^2\right),
\ee 
where $C_1$ and $C_2$ are constants of integration representing
the two degrees of freedom in our reduced system. These two 
degrees of freedom can be represented physically by the initial velocity 
and displacement of a perturbed fluid particle in the radial direction. 
The radial Lagrangian displacement $\xi_x$ is obtained from 
equation (\ref{SOLVX}) by direct integration,\footnote{In our notation, 
a subscript $x$ or $y$ on the symbol $\xi$ indicates a Lagrangian displacement, not 
a component of the potential vorticity, which is a scalar.}
\be\label{SOLX}
\xi_x = \int \delta v_x \, dt = -\frac{C_2}
{\qe \Omega {\rm Ri}} \,F\left(\frac{1 - \alpha}{4},\frac{1 + \alpha}{4};
\frac{1}{2};-\tilde{\tau}^2\right) + \frac{C_1}{\qe \Omega}\,\tilde{\tau}\,
F\left(\frac{3 - \alpha}{4},\frac{3 + \alpha}{4};\frac{3}{2};
-\tilde{\tau}^2\right).
\ee
The solutions for the other perturbation variables can be obtained from 
equations (\ref{LIN1}), (\ref{LIN5}) and (\ref{DH}) with $k_z = 0$:
\be\label{XIY}
\delta v_y = -\tilde{\tau} \delta v_x,
\ee
\be\label{SOX}
\frac{\delta \Sigma}{\Sigma_0} = \frac{\xi_x}{L_S}
\ee
and
\be\label{SOLDH}
\frac{\delta P}{P_0} = \frac{\gamma \Omega}{i c_s k_y} \left[\qe 
\tilde{\tau} \dv{}{\tilde{\tau}}\left(\frac{\delta v_x}{c_s}\right) + 
2(\qe - 1) \frac{\delta v_x}{c_s}\right].
\ee
It can be seen from the latter equation and the solution for $\delta
v_x$ that $\delta P/P_0$ remains small compared to $\delta v_x/c_s$ in the
short-wavelength limit. This demonstrates the consistency of the Boussinesq approximation. 

The hypergeometric functions can be transformed to a form valid for
large $\tilde{\tau}$ (see \citealt{as72} equations 15.3.7 and 15.1.1).
An equivalent form of the solution for $|\tilde{\tau}| \gg 1$ is
\begin{eqnarray}
\delta v_x = (C_1 V_1 + {\rm{sgn}}(\tilde{\tau}) C_2 V_2) \, |\tilde{\tau}|^
{\frac{\alpha - 3}{2}} F\left(\frac{3 - \alpha}{4}, \frac{5 - \alpha}{4}; 1 -
 \frac{\alpha}{2}; -\frac{1}{\tilde{\tau}^2}\right) + \nonumber \\ (C_1 V_3 +
 {\rm{sgn}}(\tilde{\tau}) C_2 V_4) \, |\tilde{\tau}|^{-\frac{\alpha+3}{2}} 
F\left(\frac{3 + \alpha}{4}, \frac{5 + \alpha}{4}; 1 + \frac{\alpha}{2}; 
-\frac{1}{\tilde{\tau}^2}\right), \; \; \;
\end{eqnarray}
where ${\rm{sgn}}(\tilde{\tau})$ is the arithmetic sign of
$\tilde{\tau}$ and the constants $V_i$ are given by
\begin{eqnarray}
V_1 \equiv \frac{\Gamma\left(\frac{1}{2}\right) \Gamma\left(\frac{\alpha}
{2}\right)} {\Gamma\left(\frac{3 + \alpha}{4}\right)\Gamma\left(-\frac{1 -
 \alpha}{4}\right)} \;\; , \;\; V_2 \equiv \frac{\Gamma\left(\frac{3}{2}\right)
 \Gamma\left(\frac{\alpha}{2}\right)} {\Gamma\left(\frac{5 + \alpha
 }{4}\right)\Gamma\left(\frac{1 + \alpha}{4}\right)} , \;\; \nonumber \\
V_3 \equiv \frac{\Gamma\left(\frac{1}{2}\right) \Gamma\left(-\frac{\alpha}
{2}\right)} {\Gamma\left(\frac{3 - \alpha}{4}\right)\Gamma\left(-\frac{1 +
 \alpha}{4}\right)} \;\; , \;\; V_4 \equiv \frac{\Gamma\left(\frac{3}{2}\right)
 \Gamma\left(-\frac{\alpha}{2}\right)} {\Gamma\left(\frac{5 - \alpha}
{4}\right)\Gamma\left(\frac{1 - \alpha}{4}\right)}.
\end{eqnarray}
Expanding the above form of the solution for $|\tilde{\tau}| \gg 1$, we obtain
\be\label{DVAS}
\delta v_x = \left(C_1 V_1 + {\rm{sgn}}(\tilde{\tau}) C_2 V_2\right)\,
 |\tilde{\tau}|^{\frac{\alpha - 3}{2}} + \left(C_1 V_3 + {\rm{sgn}}
(\tilde{\tau}) C_2 V_4\right) \, |\tilde{\tau}|^{-\frac{\alpha + 3}{2}} + 
O(\tilde{\tau}^{-2}).
\ee

An equivalent form of $\xi_x$ for $|\tilde{\tau}| \gg 1$ is
\begin{eqnarray}
\xi_x = \left(-\frac{C_2 X_1}{\qe \Omega {\rm Ri}} + {\rm{sgn}}
(\tilde{\tau}) \frac{C_1 X_2}{\qe \Omega}\right) \, |\tilde{\tau}|^{\frac{
\alpha - 1}{2}} F\left(\frac{3 - \alpha}{4}, \frac{1 - \alpha}{4}; 1 - 
\frac{\alpha}{2}; -\frac{1}{\tilde{\tau}^2}\right) + \nonumber \\ \left(
-\frac{C_2 X_3}{\qe \Omega {\rm Ri}} + {\rm{sgn}}(\tilde{\tau}) 
\frac{C_1 X_4}{\qe \Omega}\right) \, |\tilde{\tau}|^{-\frac{\alpha+1}{2}} 
F\left(\frac{3 + \alpha}{4}, \frac{1 + \alpha}{4}; 1 + \frac{\alpha}{2}; 
-\frac{1}{\tilde{\tau}^2}\right), \; \; \;
\end{eqnarray}
where the constants $X_i$ are given by
\begin{eqnarray}
X_1 \equiv \frac{\Gamma\left(\frac{1}{2}\right) \Gamma\left(\frac{\alpha}
{2}\right)} {\Gamma\left(\frac{1 + \alpha}{4}\right)\Gamma\left(\frac{1 + 
\alpha}{4}\right)} \;\; , \;\; X_2 \equiv \frac{\Gamma\left(\frac{3}{2}\right)
 \Gamma\left(\frac{\alpha}{2}\right)} {\Gamma\left(\frac{3 + \alpha}
{4}\right)\Gamma\left(\frac{3 + \alpha}{4}\right)} , \;\; \nonumber \\
X_3 \equiv \frac{\Gamma\left(\frac{1}{2}\right) \Gamma\left(-\frac{\alpha}
{2}\right)} {\Gamma\left(\frac{1 - \alpha}{4}\right)\Gamma\left(\frac{1 - 
\alpha}{4}\right)} \;\; , \;\; X_4 \equiv \frac{\Gamma\left(\frac{3}{2}\right)
 \Gamma\left(-\frac{\alpha}{2}\right)} {\Gamma\left(\frac{3 - \alpha}
{4}\right)\Gamma\left(\frac{3 - \alpha}{4}\right)}.
\end{eqnarray}
Expanding $\xi_x$ for $|\tilde{\tau}| \gg 1$ yields
\be\label{DSAS}
\xi_x = \left(-\frac{C_2 X_1}{\qe \Omega {\rm Ri}} + {\rm{sgn}}
(\tilde{\tau}) \frac{C_1 X_2}{\qe \Omega}\right) \, |\tilde{\tau}|^
{\frac{\alpha - 1}{2}}  + \left(-\frac{C_2 X_3}{\qe \Omega {\rm Ri}} + 
{\rm{sgn}}(\tilde{\tau}) \frac{C_1 X_4}{\qe \Omega}\right) \, |\tilde{\tau}|^
{-\frac{\alpha+1}{2}} + O(\tilde{\tau}^{-2}). 
\ee

The dominant contribution for each perturbation variable at late times is thus
\be
\delta P \propto \delta v_x \sim t^{\frac{\alpha - 3}{2}},
\ee
\be
\delta \Sigma \propto \xi_x \sim t^{\frac{\alpha - 1}{2}},
\ee
and
\be
\delta v_y \propto t \delta v_x \sim t^{\frac{\alpha - 1}{2}}.
\ee
This leads to one of our main conclusions: the density and $y$-velocity
perturbations will grow asymptotically for $\alpha > 1$, i.e.
${\rm{Ri}} \propto N_x^2 < 0$.\footnote{Notice that this is the same time 
dependence obtained by \cite{ehr53} in a modal analysis; see the discussion 
surrounding equation (\ref{ALPHA}). These power law time-dependences 
can be obtained more efficiently by solving the large-$\tilde{\tau}$ limit 
of equation (\ref{SOLVX}).} For small Richardson number, 
however (as is expected for a Keplerian disk with modest radial gradients), 
$\alpha \sim 1 - 2{\rm Ri}$ and the asymptotic growth is extremely slow:
\be
\delta \Sigma \sim \delta v_y \sim t^{-{\rm Ri}}.
\ee

In the stratified shearing sheet, the right-hand side of equation 
(\ref{PVEVLIN}) governing the evolution of the perturbed potential 
vorticity is no longer zero. The form of this equation for the incompressive 
shwaves is
\be\label{DPVEV}
\dot{\delta \xi} = {d\over{d t}}\left({i\tilde{k}_x \delta v_y - i k_y \delta v_x
\over{\Sigma_0}}\right) = \frac{c_s^2 k_y}{i L_P \Sigma_0^2}\delta \Sigma.
\ee
The asymptotic time dependence of the perturbed potential vorticity can
be obtained by integrating equation (\ref{DPVEV}):
\be
\delta \xi \sim t^{\frac{\alpha + 1}{2}} \sim t^{1 - {\rm Ri}}
\ee
for $\tilde{\tau} \gg 1$ and $|{\rm Ri}| \ll 1$. As noted in \S 2, an entropy 
gradient is not required to generate vorticity. For $N_x^2 = 0$, $\alpha = 1$ 
and the perturbed potential vorticity grows linearly with time. The unstratified 
shearing sheet is recovered in the limit of zero stratification ($1/L_P \rightarrow 
0$), since in this limit equation (\ref{DPVEV}) reduces to $\xi = constant$.

\subsection{Energetics of the Incompressive Shwaves}

For a physical interpretation of the incompressive shwaves in the stratified 
shearing sheet, we repeat the analysis of section 3.3 for the solution 
given in the previous section. For a complete description of the energy in 
this case, however, we must include the potential energy of a fluid element 
displaced in the radial direction. Following \cite{jwm61}, an  
expression for the energy in the Boussinesq approximation is obtained by
summing equation (\ref{LIN2}) multiplied by $\delta v_x$ 
and equation (\ref{LIN3}) multiplied by $\delta v_y$. Replacing 
$\delta \Sigma/\Sigma_0$ by $\xi_x/L_S$ via equation (\ref{SOX}) 
results in the following expression for the energy evolution:
\be\label{ENERGY}
\dv{E_k}{\tilde{\tau}} \equiv \dv{}{\tilde{\tau}} \left(\frac{1}{2}\Sigma_0 
\delta v^2 + \frac{1}{2} \Sigma_0 N_x^2 \xi_x^2 \right) = \Sigma_0 \delta 
v_x \delta v_y,
\ee
where $\delta v^2 = \delta v_x^2 + \delta v_y^2$. The three terms in equation 
(\ref{ENERGY}) can be identified as the kinetic energy, potential energy 
and Reynolds stress associated with an individual shwave. One may readily 
verify that the vortical shwaves (see equations (\ref{IVX})-(\ref{IS})) in the 
unstratified shearing sheet ($N_x^2 = 0$) satisfy equation (\ref{ENERGY}).

The right hand side of equation (\ref{ENERGY}) can be rewritten $-\tilde{\tau}
\delta v_x^2$ and individual trailing shwaves ($\tilde{\tau} > 0$) are therefore
associated with a negative angular momentum flux.  If the energy were
positive definite this would require that individual shwaves always
decay.  But when $N_x^2 < 0$ (${\rm Ri} < 0$) the potential energy
associated with a displacement is negative, so the energy $E_k$ can be
negative and a negative angular momentum flux is not enough to halt
shwave growth.  

Our next step is to write the constants of integration $C_1$ and $C_2$ 
in terms of the initial radial velocity and displacement of the shearing 
wave, $\delta v_{x0}$ and $\xi_{x0}$:
\be
C_1 = \frac{\qe \Omega {\rm Ri}\, \delta v_{x2}(\tilde{\tau}_0) \, 
\xi_{x0} + \xi_{x1}(\tilde{\tau}_0) \, \delta v_{x0}}{\delta v_{x1}
(\tilde{\tau}_0) \, \xi_{x1}(\tilde{\tau}_0) + {\rm Ri}\,\delta v_{x2}
(\tilde{\tau}_0) \, \xi_{x2}(\tilde{\tau}_0)} \;\; , \;\;
C_2 = \frac{-\qe \Omega {\rm Ri}\, \delta v_{x1}(\tilde{\tau}_0) \,
\xi_{x0} + {\rm Ri}\,\xi_{x2}(\tilde{\tau}_0) \, \delta v_{x0}}{\delta 
v_{x1}(\tilde{\tau}_0) \, \xi_{x1}(\tilde{\tau}_0) + {\rm Ri}\,\delta 
v_{x2}(\tilde{\tau}_0) \, \xi_{x2}(\tilde{\tau}_0)},
\ee
where $\tilde{\tau}_0 = k_{x0}/k_y$, $\delta v_{x1}$ is the 
hypergeometric function given by equation (\ref{SOLVX}) with $C_1 
= 1$ and $C_2 = 0$, and the other functions are similarly defined. These 
expressions can be simplified by noticing that the denominator of $C_1$ 
and $C_2$ is the Wronskian of the differential equation for $\xi_x$:\footnote{
Based upon the relationship between a hypergeometric function 
and its derivatives, $\delta v_{x1} = d(\xi_{x2})/d\tilde{\tau}$ and 
${\rm Ri} \delta v_{x2} = -d(\xi_{x1})/d\tilde{\tau}$.}
\be\label{ODEX}
(1 + \tilde{\tau}^2)\dv{^2\xi_x}{\tilde{\tau}^2} + 2 \tilde{\tau} 
\dv{\xi _x}{\tilde{\tau}} + {\rm Ri}\xi_x = 0.
\ee
The Wronskian of this equation is
\be
{\cal W} \equiv \dv{\xi_{x2}}{\tilde{\tau}} \, \xi_{x1}
-  \dv{\xi_{x1}}{\tilde{\tau}} \, \xi_{x2}
= \exp \left(-\int^{\tilde{\tau}} \frac{2\tau^2}
{1 + \tau^2} \, d\tau \right) =  \frac{1}{1 + \tilde{\tau}^2}.
\ee
We further simplify the analysis by setting the initial displacement $\xi_{x0}$
to zero.

With these simplifications, the solution given by equations 
(\ref{SOLVX}) and (\ref{SOLX}) becomes
\begin{eqnarray}
\frac{\delta v_x}{\delta v_{x0}} = \left(1 + \tilde{\tau}_0^2 \right)
\left[F\left(\frac{1 - \alpha}{4},\frac{1 + \alpha}{4};\frac{1}{2};
-\tilde{\tau}_0^2\right) \,F\left(\frac{3 - \alpha}{4},\frac{3 + \alpha}{4};
\frac{1}{2};-\tilde{\tau}^2\right) + \right. \nonumber \\ \left. {\rm Ri}\,
\tilde{\tau}_0\,F\left(\frac{3 - \alpha}{4},\frac{3 + \alpha}{4};
\frac{3}{2};-\tilde{\tau}_0^2\right) \, \tilde{\tau}\,F\left(\frac{5 - \alpha}
{4},\frac{5 + \alpha}{4};\frac{3}{2};-\tilde{\tau}^2\right)\right],
\end{eqnarray}
\begin{eqnarray}
\frac{\xi_x}{\delta v_{x0}} = \left(1 + \tilde{\tau}_0^2\right)\left[
-\frac{1}{\qe \Omega} \tilde{\tau}_0\,F\left(\frac{3 - \alpha}{4},
\frac{3 + \alpha}{4};\frac{3}{2};-\tilde{\tau}_0^2\right) \,F\left(
\frac{1 - \alpha}{4},\frac{1 + \alpha}{4};\frac{1}{2};-\tilde{\tau}^2
\right) + \right. \nonumber \\ \left. \frac{1}{\qe \Omega} F\left(\frac{1 
- \alpha}{4},\frac{1 + \alpha}{4};\frac{1}{2};-\tilde{\tau}_0^2\right) 
\,\tilde{\tau}\, F\left(\frac{3 - \alpha}{4},\frac{3 + \alpha}{4};
\frac{3}{2};-\tilde{\tau}^2\right)\right].
\end{eqnarray}

As in section 3.3, the energy integral for the incompressive perturbations 
is given by
\be
\< E_i \> = \frac{1}{2}\Sigma_0 L^2 \int k_0 dk_0 \< \delta v_\perp^2(k_0)\>
\int_0^{2\pi} d\theta \sin^2\theta \left[ \left(1 + \tilde{\tau}^2\right) 
\left(\frac{\delta v_x}{\delta v_{x0}}\right)^2 + N_x^2 \left(
\frac{\xi_x}{\delta v_{x0}}\right)^2\right],
\ee
for initial perturbations perpendicular to and isotropic in $\bk_0$. 
Changing integration variables to $\tilde{\tau} = \qe\Omega t + 
\cot\theta$, the angular integral becomes
\begin{eqnarray}\label{EINT}
2 \int_{-\infty}^{\infty} d\tilde{\tau} \, 
\left[ \left(1 + \tilde{\tau}^2\right) \left\{\xi_{x1}
(\tilde{\tau} - \qe \Omega t)\delta v_{x1}(\tilde{\tau}) + {\rm{Ri}}
 \, \xi_{x2}(\tilde{\tau} - \qe \Omega t) \delta v_{x2}(\tilde{\tau})
\right\}^2 + \right. \nonumber \\ \left. {\rm Ri} \, \left\{\xi_{x2}
(\tilde{\tau} - \qe \Omega t)\xi_{x1} (\tilde{\tau}) - \xi_{x1}
(\tilde{\tau} - \qe \Omega t) \xi_{x2}(\tilde{\tau})\right\}^2\right],
\end{eqnarray}
where we have used the relation $\sin\theta = (1 + \tilde{\tau}_0^2)
^{-1}$. In the limit of large $\qe \Omega t$, the dominant 
contribution to the angular integral comes from the region $0 \lesssim 
\tilde{\tau} \lesssim \qe \Omega t$. This can be seen from the 
following argument. Using the expansions given by equations 
(\ref{DVAS}) and (\ref{DSAS}), we find the angular integrand is
\be
2 |\tilde{\tau}(\tilde{\tau} - \qe \Omega t)|^{\alpha-1}\left[(V_1 
X_1  + {\rm sgn}(\tilde{\tau}){\rm sgn}(\tilde{\tau} - \qe \Omega 
t){\rm{Ri}} \, V_2 X_2)^2 + {\rm Ri} X_1^2 X_2^2 ({\rm sgn}
(\tilde{\tau})  - {\rm sgn}(\tilde{\tau} - \qe \Omega t))^2\right]
\ee
for $|\tilde{\tau}| \gg 1$ and $|\tilde{\tau} - \qe \Omega t| \gg 1$. 
Using the relation $\Gamma(n+1) = n\Gamma(n)$, one can easily
show that
\be
X_2 = \frac{2}{\alpha - 1}V_1 \;\;\;\; {\rm and} \;\;\;\; V_2 = \frac{2}
{\alpha + 1}X_1.
\ee
The integrand therefore simplifies to
\be
|\tilde{\tau}(\tilde{\tau} - \qe \Omega t)|^{\alpha-1}V_1^2 
X_1^2\frac{2}{1 - \alpha}\left[{\rm sgn}
(\tilde{\tau}) - {\rm sgn}(\tilde{\tau} - \qe \Omega t)\right]^2,
\ee
which is zero unless $0 < \tilde{\tau} < \qe \Omega t$ (for $t > 0$). 
For large $\qe \Omega t$, therefore, the angular integral is 
approximately given by
\be
\frac{16 V_1^2 X_1^2}{1 - \alpha} \int_\nu^{\qe \Omega 
t - \nu} d\tilde{\tau} \, \left[\tilde{\tau}\left(\tilde{\tau}-\qe 
\Omega t\right)\right]^{\alpha-1} = \frac{16 V_1^2 X_1^2}{\alpha 
(1 - \alpha)} \left. \frac{(\tilde{\tau} \qe \Omega t)^{\alpha}}
{\qe \Omega t} F\left(\alpha,1-\alpha;1+\alpha;\frac{\tilde
{\tau}}{\qe \Omega t}\right) \right | ^{\qe \Omega t - \nu}_{\nu},
\ee
where $1 \ll \nu \ll  \qe \Omega t$. For $\qe \Omega t \gg \nu$, the 
above expression can be approximated by evaluating it at 
$\tilde{\tau} = \qe \Omega t$, giving
\be
\< E_i (\qe \Omega t \gg 1) \>  \simeq \, 16 V_1^2 X_1^2
\frac{\Gamma(1+\alpha) \Gamma(\alpha)}{\alpha (1 - \alpha) 
\Gamma(2\alpha)} (\qe \Omega t)^{2\alpha-1} \, \< E_i(t = 0) \>,
\ee
where we have used equation 15.3.7 in \cite{as72} to evaluate 
$F(a,b;c;1)$.\footnote{We have numerically integrated the angular 
integral (\ref{EINT}) and found this to be an excellent approximation 
at late times.}

Notice that there is no power-law growth in the perturbation energy for
${\rm{Ri}} > 1/4$,\footnote{For ${\rm{Ri}} > 1/4$, $\alpha$ is 
imaginary and ${\rm{Re}}[t^{2\alpha-1}] = t^{-1}\cos(2|\alpha| \ln t)$.}
consistent with the classical Richardson criterion (\ref{RICH}). In 
our analysis the energy decays with time for $2\alpha-1<0$, or 
${\rm{Ri}} > 3/16$.  Thus the energy of an initial isotropic set of 
incompressive perturbations in a radially-stratified shearing sheet-model 
grows asymptotically (for ${\rm{Ri}} < 3/16$), just like the compressive 
shwaves and {\it unlike} the incompressive shwaves in an unstratified 
shearing sheet, for which the energy is constant in time.

The growth of an ensemble of incompressive shwaves in a stratified disk is {\it not}
due to a Rayleigh-Taylor or convective type instability.  There is 
asymptotic growth for $0 < {\rm{Ri}} < 3/16$, and convective 
instability requires ${\rm{Ri}} < 0$.  One can also see 
this by examining the asymptotic energy for small values of
$|{\rm{Ri}}|$, such as would be expected for a Keplerian disk with
modest radial gradients:
\be
\< E_i (\qe \Omega t \gg 1) \>  \simeq \left[2 \pi^2 {\rm Ri} + O({\rm Ri}^2)\right]\qe 
\Omega t^{1-4{\rm Ri}+O({\rm Ri}^2)} \, \< E_i(t = 0) \>.
\ee
Evidently for small values of Ri the near-linear growth in time of the
energy is independent of the sign of Ri and therefore $N_x^2$.\footnote{
This asymptotic expression assumes $\Ri \neq 0$. Notice that the energy 
at late times can have the opposite sign to the initial energy because the 
potential energy is negative for $N_x^2 < 0$.}

\section{Implications}

We have studied the nonaxisymmetric linear theory of a thin, 
radially-stratified disk.  Our findings are: (i) incompressive,
short-wavelength perturbations in the unstratified shearing sheet
exhibit transient growth and asymptotic decay, but the energy of an
ensemble of such shwaves is constant with time (consistent with \citealt
{amn04}); (ii) short-wavelength compressive shwaves grow asymptotically
in the unstratified shearing sheet, as does the energy of an ensemble of
such shwaves, which in the absence of any other dissipative effects
(e.g., radiative damping) will result in a compressive shwave steepening
into a train of weak shocks; (iii) incompressive shwaves in the
stratified shearing sheet have density and azimuthal velocity
perturbations $\delta \Sigma$, $\delta v_y \sim t^{-{\rm Ri}}$ (for
$|{\rm Ri}| \ll 1$); (iv) incompressive shwaves in the stratified
shearing sheet are associated with an angular momentum flux proportional
to $-\tilde{k}_x/k_y$; leading shwaves therefore have positive angular momentum
flux and trailing shwaves have negative angular momentum flux; (v) the
energy of an ensemble of incompressive shwaves in the stratified
shearing sheet behaves asymptotically as $t^{1-4{\rm Ri}}$ for $|{\rm
Ri}| \ll 1$.  For Keplerian disks with modest radial gradients, $|{\rm
Ri}|$ is expected to be $\ll 1$, and there will therefore be weak growth
in a single shwave for ${\rm Ri} < 0$ and near-linear growth in the
energy of an ensemble of shwaves, independent of the sign of Ri.

Along the way we have found the following solutions: (i) an exact solution 
for non-vortical shwaves in the unstratified shearing sheet, equations 
(\ref{CVX}), (\ref{CS}) and (\ref{CVY}); (ii) a WKB-solution for the 
non-vortical,
compressive shwaves in the short-wavelength, high-frequency limit, 
equations (\ref{VYWKB})-(\ref{SWKB}); (iii) 
a solution for incompressive shwaves in the unstratified shearing sheet 
valid in the short-wavelength, low-frequency limit, equations 
(\ref{IVX})-(\ref{IS}); (iv) a solution for incompressive shwaves in 
the radially-stratified shearing sheet (also valid in the short-wavelength, 
low-frequency limit), equations (\ref{SOLVX})-(\ref{SOLDH}). 

Our results are summarized in Figure 1, which shows the regions of
amplification and decay for shwaves in a stratified disk in the
$N_x^2/\Omega, \tilde{q}$ plane.  

The presence of power-law growth of incompressive shwaves in stratified
disks opens the possibility of a transition to turbulence as amplified
shwaves enter the nonlinear regime.  Any such transition would depend,
however, on the nonlinear behavior of the disk after the shwaves break.
It is far from clear that they would continue to grow.  We will evaluate
the nonlinear behavior of the disk in subsequent work.

Our results are essentially in agreement with the numerical results
presented by \cite{klr04} (although the decay of high frequency,
compressive shwaves seen in his Figures 8, 9, 10 is likely a numerical
effect since compressive shwaves grow asymptotically in our analysis),
that is, we find that arbitrarily large amplification factors can be
obtained by starting with appropriate initial conditions.  Our results,
however, clarify the nature and asymptotic time dependence of the
growth.  Our results on the unstratified shearing sheet are also
consistent with the results of \cite{amn04}, who find that an isotropic
ensemble of incompressive shwaves have fixed energy.

This work was supported by NSF grant AST 00-03091 and a Drickamer
Fellowship for BMJ. We thank Jeremy Goodman for his comments.

\newpage

\begin{appendix}

We demonstrate here that the Boussinesq approximation to the linear 
perturbation equations is formally equivalent to a short-wavelength, 
low-frequency limit of the full set of linear equations. We perform the 
demonstration for the stratified shearing-sheet model since the standard 
shearing sheet is recovered in the limit of zero stratification.

Combining equations (\ref{LIN1a}) through (\ref{LIN5a}) into a single 
equation for $\delta v_x$ yields the following differential equation, 
fourth-order in time:
\be\label{DV4DT}
F_4 \delta v_x^{(4)} + F_3 \delta v_x^{(3)} + F_2\delta v_x^{(2)}  + F_1 
\delta v_x^{(1)}  + F_0 \delta v_x = 0,
\ee
where
\be\label{COEF4}
F_4 = \tilde{k}_x^2 \left[(k_y^2 + k_z^2) \left(1-\frac{i}{\tilde{k}_x L_P}
\right)^2 + k_y^2\frac{2(\qe+1)(\qe+2)}{\tilde{k}_x^2 H^2}\right],
\ee
\be
F_3 = -2 \qe \Omega \tilde{k}_x k_y (k_y^2 + k_z^2) \left(1+\frac{i}
{\tilde{k}_x L_P}\right),
\ee
\begin{eqnarray}
F_2 = c_s^2 \tilde{k}_x^2 \left[\tilde{k}_x^2 \left(1 + \frac{1}{\tilde{k}
_x^2 L_P^2} + \frac{N_x^2 + \tilde{\kappa}^2}{\tilde{k}_x^2 H^2}\right) 
\left\{ (k_y^2 + k_z^2)\left(1-\frac{i}{\tilde{k}_x L_P}\right)^2 +  k_y^2 
\frac{2(\qe + 1)(\qe + 2)}{\tilde{k}_x^2 H^2} \right\} \, + \right. \nonumber 
\\ \left. (k_y^2 + k_z^2) \left\{(k_y^2 + k_z^2)\left(1-\frac{i}{\tilde{k}_x L_P}
\right)^2 + k_y^2 \frac{2(\qe + 1)(3\qe + 2)}{\tilde{k}_x^2 H^2}\right\} \right],
\end{eqnarray}
\begin{eqnarray}
F_1 = 4\qe \Omega c_s^2 k_y \tilde{k}_x^3\left[(k_y^2 + k_z^2)\left(1+\frac{i}{\tilde{k}_x L_P}\right) \left\{1 + \frac{i(3\qe - 2)}
{2\qe \tilde{k}_x L_P} + \frac{\tilde{\kappa}^2}{4\qe \tilde{k}_x^2 L_P^2} 
- \frac{N_x^2 + \tilde{\kappa}^2}{2\tilde{k}_x^2 H^2}\right\} \, + \right. 
\nonumber \\ \left. 3k_y^2\frac{(\qe + 1)(\qe + 2)}{\tilde{k}_x^2 H^2}
\left(1 - \frac{2i}{3\qe \tilde{k}_x L_P}\right) \right],
\end{eqnarray}
\be\label{COEF0}
F_0 = c_s^2 \tilde{k}_x^2 \left[k_y^2 (N_x^2 + 2\qe^2 \Omega^2) + 
k_z^2(N_x^2 + \tilde{\kappa}^2)\right] \left[(k_y^2 + k_z^2)\left(1-\frac{i}
{\tilde{k}_x L_P}\right)^2 + k_y^2 \frac{2(\qe+1)(3\qe+2)}
{\tilde{k}_x^2 H^2}\right].
\ee

The above expressions have been written to make the short-wavelength
limit more apparent: all but the leading-order terms in brackets are
proportional to factors of $(\tilde{k}_xL_P)^{-1}$ or
$(\tilde{k}_xH)^{-1}$. Notice also that since one expects $H/L_P \ll 1$ for 
Keplerian disks with modest radial gradients,
\be
\frac{1}{\tilde{k}_x L_P} = \frac{1}{k_y H\tilde{\tau}}\frac{H}{L_P} 
\ll \frac{1}{k_y H\tilde{\tau}},
\ee
(for $\tilde{\tau} \neq 0$) and therefore the short-wavelength limit is sufficient. 
One needs to be careful in taking this limit, however, since $\tilde{k}_x = k_y
\tilde{\tau}$ goes through zero as a shwave goes from leading to trailing. The 
approximation is rigorously valid only for $\tilde{\tau} \neq 0$, but we have 
numerically integrated the full set of linear equations (equations (\ref{LIN1a}) 
through (\ref{LIN5a}) with $k_z = 0$) and found good agreement with the 
Boussinesq solutions described in \S 4.3 for all $\tilde{\tau}$ at sufficiently 
short wavelengths.\footnote{One must start with a set of initial conditions 
consistent with equations (\ref{LIN2}), (\ref{LIN3}), (\ref{LIN1}) and 
(\ref{LIN5}) in order to accurately track the incompressive-shwave solutions. 
In addition, suppression of the high-frequency compressive-shwave solutions 
near $\tilde{\tau} = 0$ requires $k_y L_P \gtrsim 200$, which for $H/L_P = 
0.1$ implies $H k_y \gtrsim 20$.}

With these assumptions in mind, to leading order in $(H k_y)^{-1}$ equation 
(\ref{DV4DT}) becomes
\begin{eqnarray}\label{EQVX}
\tilde{k}_x \delta v_x^{(4)} - 2 \qe \Omega k_y \delta v_x^{(3)} + 
c_s^2 \tilde{k}_x \tilde{k}^2 \delta v_x^{(2)}  + 4\qe \Omega c_s^2 
\tilde{k}_x^2 k_y \delta v_x^{(1)} \, + \nonumber \\ c_s^2 \tilde{k}_x 
\left[k_y^2(N_x^2 + 2\qe^2 \Omega^2) + k_z^2(N_x^2 + \tilde{\kappa}^2)
\right]\delta v_x = 0.
\end{eqnarray}
If we assume $\partial_t \ll c_s k_y$, the two highest-order time derivatives 
are of lower order and can be neglected (thereby eliminating the compressive 
shwaves) and we have
\be\label{DV2DT}
\tilde{k}^2 \ddot{\delta v_x} + 4 \qe \Omega \tilde{k}_x k_y\dot{\delta v_x}
\left[k_y^2(N_x^2 + 2\qe^2 \Omega^2) + k_z^2(N_x^2 +\tilde{\kappa}^2)\right] 
\delta v_x = 0.
\ee
This is equivalent to equation (\ref{BOUSSVX}).

Notice also that the assumption $\partial_t \sim O(c_s k_y)$ applied to 
equation (\ref{EQVX}) yields
\be
\delta v_x^{(4)} + c_s^2 \tilde{k}^2 \delta v_x^{(2)} = 0
\ee
to leading order in $(H k_y)^{-1}$. This equation is of the same form 
as the short-wavelength limit of equation (\ref{SOUND}) for the 
compressive shwaves in the unstratified shearing sheet, confirming our 
claim that short-wavelength compressive shwaves are unchanged at 
leading order by stratification.

\end{appendix}

\begin{figure}
\plotone{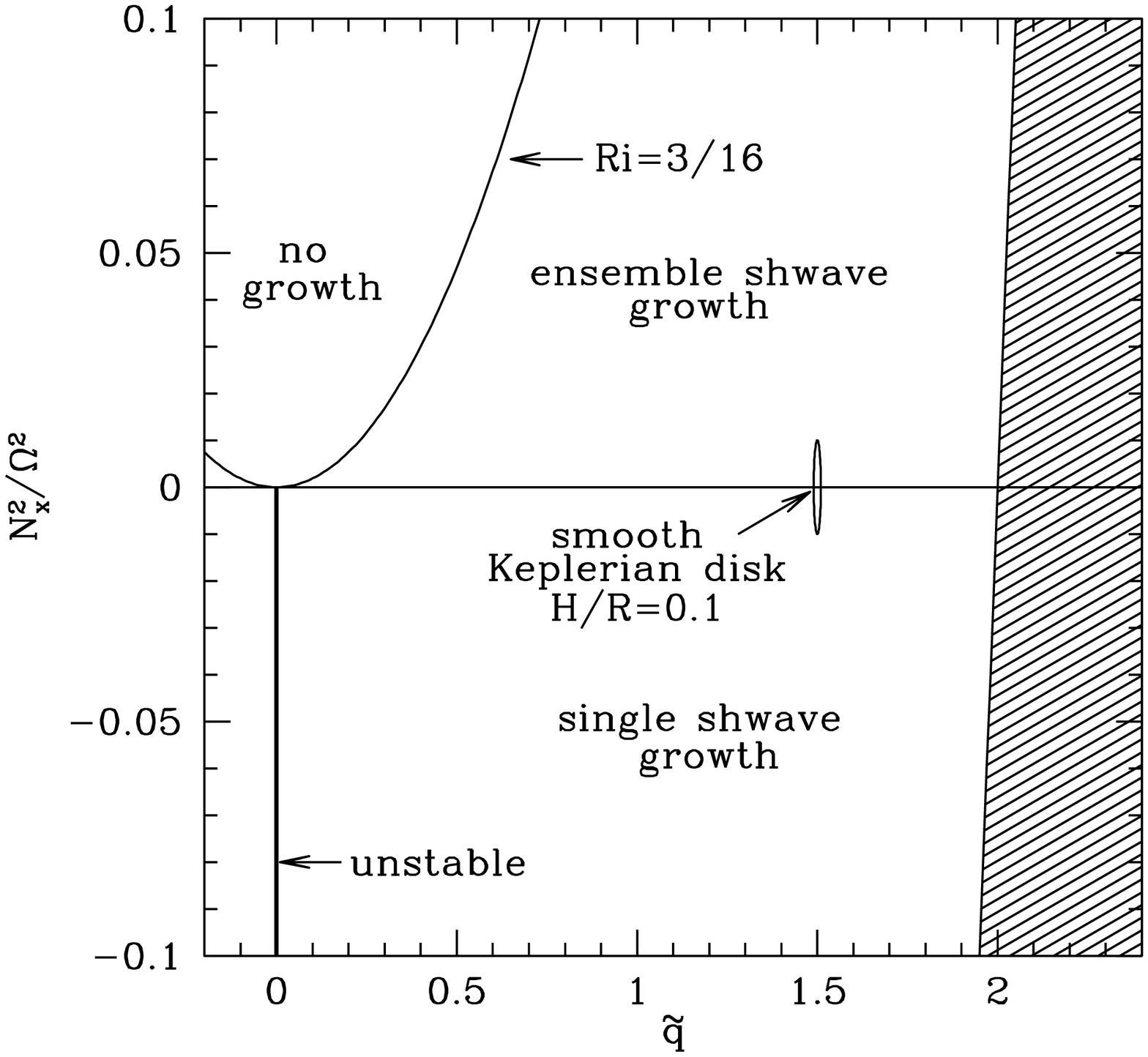}
\caption{
A summary of analytic results for shwaves (shearing waves) in a
stratified disk.  The relevant parameters are the local shearing rate
$\tilde{q} = -\frac{1}{2}d\ln\Omega^2/d\ln r$ and the dimensionless
Brunt-V\"ais\"al\"a frequency $N_x^2/\Omega^2$.  The expected location
of a thin, smooth disk is shown as a vertically extended ellipse near
$\tilde{q} = 0$, $N_x^2/\Omega^2 = 0$.  The far right region (shaded in
the figure) is forbidden by the H{\o}iland criterion.  When $\tilde{q} =
0$ shear is absent and a modal analysis is possible; instability is
present for $N_x^2 < 0$.  Solitary shwaves with $\Ri =
N_x^2/(\tilde{q}^2\Omega^2) < 0$ experience asymptotic power-law growth
($\propto t^{-\Ri}$ for small $\Ri$); since each shwave grows the energy
of an ensemble of shwaves does as well.  For $0 < \Ri < 3/16$ solitary
shwaves decay but the energy of an ensemble of shwaves grows as a
power-law in time.  For $\Ri > 3/16$ both solitary shwaves and the energy
of an ensemble of shwaves asymptotically decay.
}
\end{figure}

\end{document}